\documentclass[conference]{IEEEtran}
\IEEEoverridecommandlockouts
\usepackage{cite}
\usepackage{amsmath,amssymb,amsfonts}
\usepackage{algorithmic}
\usepackage{graphicx}
\usepackage{textcomp}
\usepackage{xcolor}
\usepackage{url}
\usepackage{booktabs}

\def\BibTeX{{\rm B\kern-.05em{\sc i\kern-.025em b}\kern-.08em
    T\kern-.1667em\lower.7ex\hbox{E}\kern-.125emX}}

\makeatletter
\newcommand{\linebreakand}{%
  \end{@IEEEauthorhalign}
  \hfill\mbox{}\par
  \mbox{}\hfill\begin{@IEEEauthorhalign}
}
\makeatother

\begin{document}

\title{ShieldShare: Building a VPN-backed Android Hotspot for Secure Internet Sharing with Per-User Traffic Accounting}

\author{\IEEEauthorblockN{1\textsuperscript{st} Carlos Semeho Edorh}
\IEEEauthorblockA{\textit{Khoury College of Computer Sciences} \\
\textit{Northeastern University}\\
Vancouver, Canada \\
edorh.c@northeastern.edu}
\and
\IEEEauthorblockN{2\textsuperscript{nd} Jialu Bi}
\IEEEauthorblockA{\textit{Khoury College of Computer Sciences} \\
\textit{Northeastern University}\\
Vancouver, Canada \\
bi.jial@northeastern.edu}
\and
\IEEEauthorblockN{3\textsuperscript{rd} Hanchen Ye}
\IEEEauthorblockA{\textit{Khoury College of Computer Sciences} \\
\textit{Northeastern University}\\
Vancouver, Canada \\
ye.hanc@northeastern.edu}

\linebreakand
 
\IEEEauthorblockN{4\textsuperscript{th} Dawood Sajjadi}
\IEEEauthorblockA{\textit{} 
\textit{IEEE Senior Member}\\
Vancouver, Canada \\
dawood.sajjadi@ieee.org}
\and
\IEEEauthorblockN{5\textsuperscript{th}Maryam Tanha}
\IEEEauthorblockA{\textit{Khoury College of Computer Sciences} \\
\textit{Northeastern University}\\
Vancouver, Canada \\
m.tanha@northeastern.edu}
 }

\maketitle

\begin{abstract}

Virtual Private Networks (VPNs) have become essential privacy tools for mobile users, yet current implementations face significant limitations in shared environments. Mainstream VPN providers impose device limits, while Android's native hotspot functionality lacks support for routing shared traffic through VPN connections. Existing solutions either require root access or lack comprehensive monitoring capabilities. This paper presents ShieldShare, a proxy-based Android application that enables secure VPN-backed hotspot sharing with per-user traffic accounting without requiring root access. Our system employs a modular architecture comprising VPN detection, hotspot management, proxy-based traffic forwarding supporting HTTP, HTTPS, and SOCKS5, and comprehensive traffic metering with quota management. Our evaluation shows that ShieldShare reliably routes client traffic through VPN tunnels while maintaining accurate per-client bandwidth allocation and accounting. This enables affordable, community-controlled secure access in censored or high-surveillance environments, benefiting activists, investigative journalists, and shared household networks. We release ShieldShare as open-source software to support further research and real-world deployment.
\end{abstract}


\section{Introduction}
 Virtual Private Networks (VPNs) are widely adopted by mobile users who want privacy, security or access to geo-restricted content. However, current VPN implementations face significant limitations for shared usage. Mainstream providers usually impose device limits. NordVPN, for example, sets a limit of 10 concurrent connections per account \cite{nordvpn-device-limit}. On the other hand, Android’s native hotspot functionality lacks support for routing shared traffic through VPN or per-user accounting. These limitations create barriers in collaborative environments such as classrooms, study groups, or small teams.

VPN-sharing applications are particularly valuable in censored or high-surveillance regions, where multiple users can tunnel traffic through an encrypted remote endpoint to access blocked content. In such contexts, VPNs are widely recommended by digital-rights and press-freedom organizations as basic security infrastructure for investigative journalists, helping them protect their sources, research, and communication patterns~\cite{digi-safety}\cite{censorship}. A VPN-sharing application can extend these benefits in a more affordable and controllable way by allowing multiple users to share a single secure VPN connection. That is, instead of requiring every individual to purchase and configure their own commercial VPN, a single well-provisioned endpoint can securely share its tunneled connection with others in the local network. This model lowers the per-user cost of circumvention, which is critical in low-income or highly censored regions where subscriptions, payment methods, or app-store access are restricted~\cite{saropa}. Existing solutions either require root access (VPNHotSpot~\cite{Mygod2025VPNHotspot}) or lack per-user traffic monitoring (Every Proxy~\cite{everyproxy}). While prior work has studied VPN security, usability, and Android constraints extensively, none systematically addresses sharing VPN connections via hotspot without root access while providing per-user traffic accounting.

Our project aims to bridge this gap by developing a one-stop Android application that: (i) detects a VPN connection on the device, (ii) enables hotspot sharing with all traffic routed through the VPN, (iii) provides per-user traffic accounting, and (iv) does not require root access. Given that users rank convenience among the highest priorities \cite{298100}, providing a comprehensive and easy-to-use solution to address the difficulty of VPN sharing is particularly valuable. Our approach is to redirect shared traffic utilizing a proxy-based method to overcome Android’s restrictions while maintaining usability. The proxy supports HTTP, HTTPS and SOCKS5 protocols, with authentication available at the user’s discretion.

Specifically, we define two research questions.
\begin{itemize}
    \item \textbf{\textit{RQ1}} How to share VPN traffic via hotspot on Android without root access?
    \item \textbf{\textit{RQ2}} How can per-user activity be fairly monitored in such environments?
\end{itemize}
By combining VPN detection, hotspot sharing, per-user monitoring and traffic management in a single Android application, this research contributes both a practical solution for secure group Internet access and the first systematic academic study of this challenge. Our research project is released as open-source software to facilitate reproducibility and encourage community collaboration\footnote{https://github.com/yestark/ShieldShare-VPN-project}.

\section{Related Work}
A VPN-sharing application can act as a cost-efficient, community-controlled layer for secure internet access, amplifying the role of VPNs as core infrastructure for resisting censorship and information control.
The closest solution to ours is \textit{Every Proxy}~\cite{everyproxy}, which is an Android app that turns a phone into a local proxy server, allowing other devices to route their traffic through the phone’s network connection without requiring root access. This makes it useful as a lightweight building block for VPN-sharing setups.
However, it is not open-source and focuses solely on proxy server functionality, lacking features useful for collaborative scenarios like per-user traffic accounting, setup assistance, and traffic management. Another relevant solution is \textit{VPNHotspot}~\cite{Mygod2025VPNHotspot}, which supports VPN-backed hotspot sharing and monitoring but requires root access.

ShieldShare addresses these gaps through an integrated, open-source solution combining VPN detection, per-user traffic metering, quota management, traffic regulation, and QR code provisioning—all within Android's non-root constraints. Our design responds to usability requirements identified by Ramesh et al.~\cite{285411} and Raj et al.~\cite{298100}, who found that speed, convenience, and user-friendly interfaces are critical for user experience. These findings motivated ShieldShare's design emphasis on integrating multiple functions.

From a security and transparency perspective, Ikram et al.~\cite{Ikram2016} revealed that 75\% of analyzed VPN applications contain tracking libraries that introduce privacy risks. ShieldShare avoids all tracking libraries and releases its complete source code publicly, enabling independent security review and addressing the transparency gap documented in prior work~\cite{285411, Ikram2016}.

\section{Design and Architecture of ShieldShare}
Traffic from devices connected to a mobile hotspot on Android bypasses the host's VPN tunnel by default. While rooting could provide a direct solution, it introduces security risks inappropriate for typical users. ShieldShare employs a proxy-based architecture: an intermediary proxy shares the VPN connection with hotspot clients, providing VPN coverage to multiple devices without root access.

Our design assumes clients trust the host device similar to trusting a standard Wi-Fi access point (AP). The proxy collects and forwards all client traffic, requiring clients to trust the host to handle their data securely. Although this model is incompatible with zero-trust architectures due to its reliance on implicit trust in shared access points, it proves effective in two key scenarios: circumventing Internet censorship, where users in regions with restricted access can obtain blocked content through a trusted intermediary sharing VPN connectivity, and enabling temporary secure access to trusted networks, such as corporate or private intranets, whereby a host with legitimate VPN privileges grants limited access to others.

\begin{figure*}[h]
    \centering
    \includegraphics[width=0.7\linewidth]{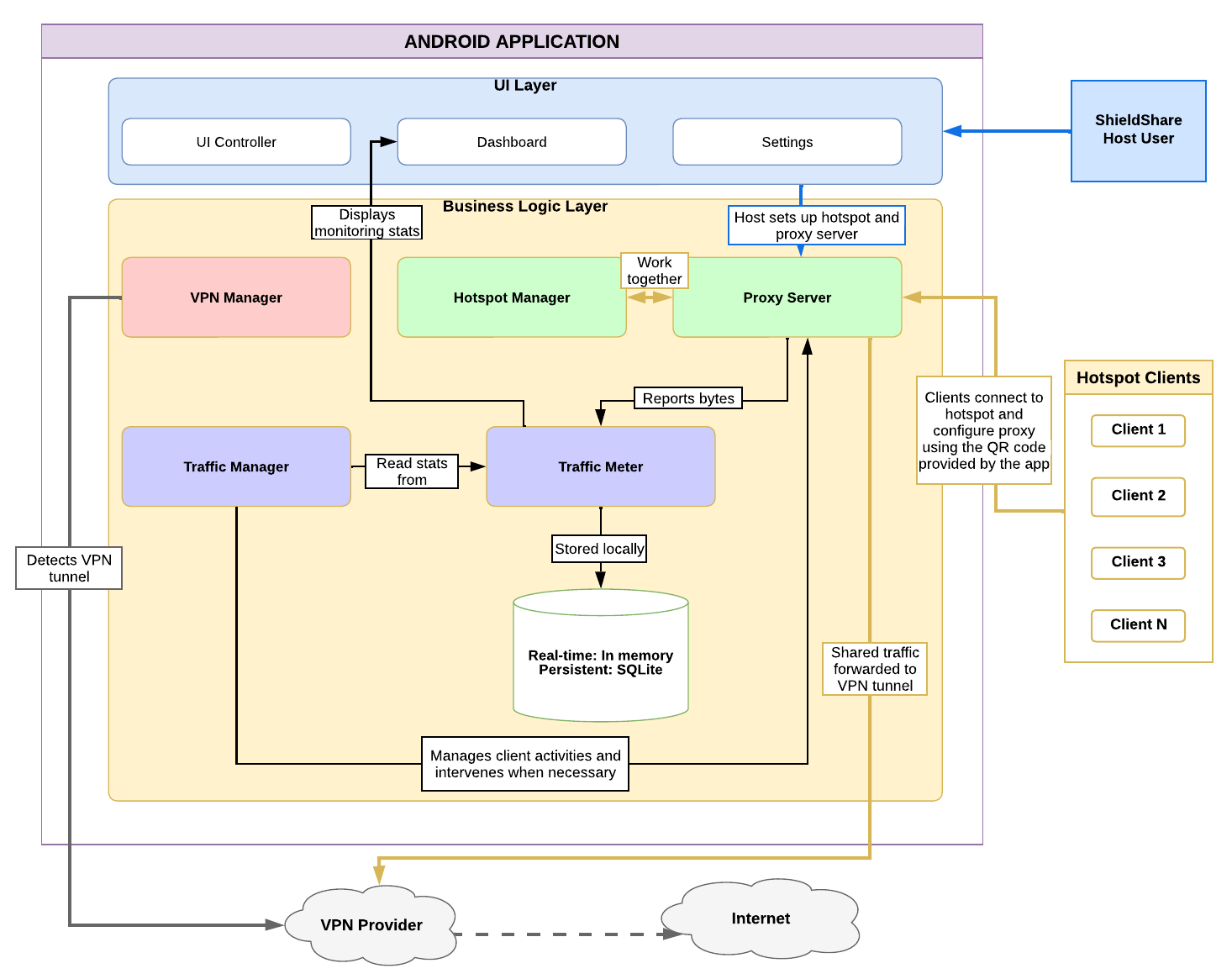}
    \caption{ShieldShare's three-layer structure: UI Layer with Dashboard and Settings at top; Business Logic Layer in middle with VPN Manager, Hotspot Manager, Proxy Server, Traffic Meter, and Traffic Manager components connected by arrows showing data flow; and storage layer with SQLite database at bottom. A host user figure on right connects to UI, multiple client figures on right connect to hotspot. VPN Provider and Internet clouds at bottom show external connections.}
    \label{fig:architecture}
    
\end{figure*}
%ShieldShare Architecture. All client traffic routed through proxy, then forwarded to VPN tunnel. 
While ShieldShare could theoretically operate on a shared Wi-Fi access point (where both host and clients connect to the same third-party AP), we deliberately design for the hotspot scenario. In a hotspot configuration, the host controls the entire layer-2 network between client and proxy. Using a shared Wi-Fi AP would introduce a third-party LAN where the AP owner could intercept traffic before it reaches the proxy. Focusing on hotspot minimizes the attack surface and maintains a clear trust boundary.

Figure \ref{fig:architecture} illustrates the complete system architecture, showing how hotspot clients connect through the proxy server, which forwards all traffic through the VPN tunnel to the internet while simultaneously reporting usage statistics to the traffic metering component.

\subsection{VPN Manager}
The \textit{VPN Manager} component observes the device’s network status to detect active VPN connections. It exposes a simple interface that allows the rest of the system to query whether a VPN tunnel is active. This component ensures that the app remains VPN-independent while still providing awareness of VPN usage.

\subsection{Hotspot Manager}
The \textit{Hotspot Manager} operates without requiring root access, making it compliant with Android's security constraints. Since Android does not provide direct Application Programming Interface (API) to enable hotspots programmatically, the component guides users through the settings to manually ensure that it's enabled.

Due to Android's restrictions on accessing Address Resolution Protocol (ARP) tables and system logs, traditional network-level client detection methods are unreliable on modern Android versions. Instead, client detection and disconnection are handled at the proxy layer, as described in Section~\ref{traffic_meter}. The Hotspot Manager's primary responsibility is therefore to facilitate hotspot setup and coordinate with other components within Android's security constraints.

\subsection{Proxy Server}
The \textit{Proxy Server} is a critical component that enables VPN-backed hotspot sharing. It listens on a specified port and accepts HTTP, HTTPS and SOCKS5 connections from hotspot clients. ShieldShare is installed only on the host device. Client devices do not require any app installation; they simply connect to the host’s hotspot and configure in their system network settings. Once connected, all traffic is automatically routed through the host’s VPN without any further interaction with the host app. When a client sends a request, the proxy server:
\begin{itemize}
    \item Authenticates the client (optional, based on configuration).
    \item Forwards the request through the VPN tunnel.
    \item Reports the byte count (upload and download) to the Traffic Meter.
    \item Returns the response to the client.
\end{itemize}

This design directly addresses \textit{\textbf{RQ1}} by routing all client traffic through the VPN tunnel without requiring root access.

\paragraph{Client Configuration Strategy and Feasibility}
ShieldShare's proxy-based approach requires client devices to configure their system settings to route traffic through the host's proxy server. This configuration cannot be automated for two reasons: first, ShieldShare cannot remotely modify network settings on client devices due to platform security restrictions; second, automatic proxy discovery protocols such as the Web Proxy Auto-Discovery Protocol (WPAD) are not supported on Android, where ShieldShare runs.

To address these limitations, ShieldShare adopts a user-assisted and lightweight provisioning flow. The host displays a QR code containing the proxy address, port, and optional credentials, along with a link to a local help page. For systems supporting PAC files, the page exposes a \texttt{proxy.pac} URL hosted by the ShieldShare device. Because unproxied traffic would bypass the VPN and accounting components, proxy configuration is mandatory for internet access through ShieldShare. While proxy configuration is required, our user assistance feature streamlines the process and makes ShieldShare more accessible to non-technical users.

\subsection{Traffic Meter}
\label{traffic_meter}
The \textit{Traffic Meter} component provides per-user accounting by recording data consumption at the proxy layer without packet-capture tools. Proxy handlers track upload and download bytes as data flows through client and target sockets in real-time, with totals persisted to SQLite after a session ends. As a fallback method for the restriction of the ARP table, the system utilizes periodic subnet probes (e.g., pinging 192.168.x.0/24) and proxy session data to monitor clients based on IP address. Each client IP receives an auto-generated identifier for display purposes, with traffic statistics tracked independently per session.

This approach eliminates the need for packet-capture tools such as Wireshark and operates fully within the non-root constraints of Android. No traffic is intercepted and only byte counts and minimal metadata are observed. It achieves near real-time client detection typically within one polling interval and accurate per-client byte accounting.

\subsection{Traffic Manager}
The \textit{Traffic Manager} coordinates traffic-related control logic beyond raw byte accounting. While the \textit{Traffic Meter} measures consumption, the \textit{Traffic Manager} governs allocation, regulation, and interpretation through three subsystems: quota assignment, performance monitoring, and traffic regulation. These mechanisms ensure fair bandwidth distribution, protect the host from abusive workloads, and provide system insights.

\subsubsection{Quota Assignment}
\label{subsubsec:quota}
To prevent bandwidth monopolization, the Traffic Manager implements capping based on \textit{Traffic Meter} data. The system supports two modes:

\begin{itemize}
    \item \textbf{Dynamic mode}: Total bandwidth is divided equally among active clients, with allocations updating automatically as clients join or disconnect. When clients leave, remaining clients receive larger shares.
    \item \textbf{Fixed mode}: Each client receives a predetermined independent quota that remains constant regardless of connection status, providing predictable bandwidth guarantees.
\end{itemize}

In both modes, clients reaching 100\% quota are temporarily blocked, triggering host notifications and a configurable cool-down period that rejects new proxy sessions.

\subsubsection{Performance Monitoring}
The Traffic Manager includes an integrated performance monitor that samples battery level, CPU load, active connections, and throughput every five seconds. A dedicated \textit{Performance Insights} view renders live metrics, allowing users to correlate resource spikes with client behaviors.

\subsubsection{Traffic Regulation}
Beyond quota enforcement, the Traffic Manager provides policy-based traffic control. A \textit{Filter Manager} stores blocked domains and ports, consulted by proxy handlers before establishing connections. Hosts can enable presets for high-bandwidth applications or specify custom domains with automatic subdomain matching. A \textit{Consumption Tracker} flags clients whose sustained usage exceeds \textit{N} times their historical baseline. Upon detecting anomalous usage, the system alerts hosts and enables disconnection or quota capping.

\section{VPN Compatibility}
\label{sec:vpn_compatibility}
ShieldShare's proxy-based architecture requires compatibility with third-party VPN applications to route client traffic through VPN tunnels. However, Android's VPN routing model creates a fundamental constraint where VPN applications control routing rules via the \texttt{VpnService} API, and ShieldShare cannot modify these rules. The critical requirement is that local network traffic (e.g., hotspot clients connecting to the proxy at 192.168.x.x:8080) must bypass the VPN tunnel, while internet-bound traffic from the proxy must route through the VPN. This requirement is satisfied by VPN applications supporting \textit{LAN Connections}, a feature that automatically excludes local IP ranges from the VPN tunnel. When enabled, incoming local connections to the proxy bypass the VPN and reach the hotspot interface directly, while the proxy's outgoing internet connections route through the VPN tunnel using VPN-aware socket factories. When disabled, all traffic including local connections routes through the VPN tunnel, which VPN servers typically drop. Table~\ref{tab:compatibility} summarizes compatibility testing results. Most mainstream providers support LAN Connections by default or as a configurable option. Some VPN services do not support the feature at all (VPN Unlimited, Thunder VPN, Lantern), making them incompatible with ShieldShare regardless of subscription tier. Proton VPN uniquely restricts LAN Connections as a paid-only feature analysis of their source code~\cite{protonvpn-android} confirms that free users cannot enable this feature.

\begin{table}[h]
\centering
\caption{VPN compatibility}
\label{tab:compatibility}
\begin{tabular}{lcc}
\hline
VPN provider&Free plans&Paid plans\\
\hline

    Nord VPN        & N/A                   & 51 \\
    Express VPN     & N/A                   & 51 \\
    Norton VPN      & N/A                   & 51 \\
    Surfshark       & N/A                   & 51\\
    VPN Proxy Mater & 51& 51 \\
    VPN Unlimited   & 55 & 55 \\
    Proton VPN      & 55 & 51 \\
    Thunder VPN     & 55& 55 \\
    Windscribe      & 51& 51 \\
    Mullvad VPN     & N/A & 51 \\
    Lantern         & 55& 55 \\
    Clash           & N/A  & 51 \\
    V2ray           & N/A  & 51 \\
    Outline VPN     & N/A& 51 \\
    \hline
    
\end{tabular}
   
\end{table}

\section{Evaluation} 
\label{sec:evaluation}

We verify ShieldShare's core functionalities and performance characteristics through a series of systematic experiments. Our evaluation addresses two key aspects: (1) network performance under maximum load conditions, and (2) resource consumption during realistic sharing scenarios.

\subsection{Network Performance}
\label{subsec:throughput}
To evaluate ShieldShare's network performance limits, we conducted stress tests measuring throughput and latency under maximum load conditions. To account for fairness, Jain's Fairness Index (JFI)~\cite{jain1984fairness} is reported for both downloads and uploads. These tests intentionally pushed all clients to simultaneously saturate available bandwidth, representing worst-case performance unlike realistic usage with mixed traffic patterns.

\subsubsection{Test Configuration}
\label{subsubsec:throughput_config}
The evaluation used battery-powered operation with a 2.4 GHz hotspot, HTTPS proxy, and NordVPN, progressively scaling from 1 to 5 concurrent clients. Each test iteration consisted of 30 seconds download followed by 30 seconds upload. Testing 5 clients represents a substantial stress load—50\% of the typical 10-device Android hotspot limit under conditions where all clients simultaneously attempt maximum throughput, an extreme edge case unlikely in practice. Three scenarios were evaluated to assess VPN server impact and baseline performance:
\begin{itemize}
    \item ShieldShare with Canada VPN server
    \item ShieldShare with Germany VPN server  
    \item Baseline: Direct hotspot without ShieldShare or VPN
\end{itemize}

Each scenario and client-count combination was repeated five times with results averaged to reduce measurement variance. Additional tests with Proton VPN and ExpressVPN revealed only minor differences in performance compared to NordVPN, confirming that the findings hold true across leading mainstream VPN providers.

\begin{table*}[h]
\centering
\caption{Throughput and Latency by Client Count}
\label{tab:throughput_comparison}
\begin{tabular}{ccccccc}
\toprule
\textbf{No. of Clients} & \textbf{Aggregate Download (Mbps)} & \textbf{Download JFI} & \textbf{Aggregate Upload (Mbps)} & \textbf{Upload JFI} & \textbf{Latency (ms)} \\ 
\midrule
\multicolumn{6}{c}{\textbf{Canada}}                                                                                      \\
\midrule
1 & $83.84 \pm 6.30$ & N/A & $104.78 \pm 10.01$ & N/A & $18.20 \pm 3.96$ \\
3 & $81.76 \pm 4.47$ & $0.96 \pm 0.016$ & $112.43 \pm 6.21$ &  $0.91 \pm 0.027$ & $17.33 \pm 3.02$ \\
5 & $84.06 \pm 5.91$ & $0.95 \pm 0.017$ & $110.02 \pm 6.42$ & $0.93 \pm 0.036$ & $13.68 \pm 1.80$ \\
\midrule
\multicolumn{6}{c}{\textbf{Germany}}                                                                                     \\
\midrule
1 & $81.98 \pm 4.85$ & N/A & $108.48 \pm 7.14$ & N/A & $164.80 \pm 8.76$ \\
3 & $83.04 \pm 2.71$ & $0.97 \pm 0.012$ & $115.52 \pm 9.30$ &  $0.93 \pm 0.018$ & $162.73 \pm 4.25$ \\
5 & $84.20 \pm 4.71$ & $0.96 \pm 0.026$ & $110.44 \pm 3.34$ & $0.90 \pm 0.034$ & $156.88 \pm 3.11$ \\
\midrule
\multicolumn{6}{c}{\textbf{Baseline}}                                                                                   \\
\midrule
1 & $89.50 \pm 7.88$ & N/A & $127.04 \pm 8.45$ & N/A & $14.20 \pm 2.17$ \\
3 & $85.94 \pm 3.48$ & $0.99 \pm 0.013$ & $123.64 \pm 7.49$ &  $0.84 \pm 0.030$ & $12.87 \pm 2.92$ \\
5 & $89.38 \pm 6.73$ & $0.98 \pm 0.021$ & $124.94 \pm 3.05$ & $0.86 \pm 0.025$ & $13.36 \pm 3.88$ \\
\bottomrule
\end{tabular}
\end{table*}

\subsubsection{Throughput Analysis}
\begin{figure}[h]
    \centering
    \includegraphics[width=0.8\linewidth]{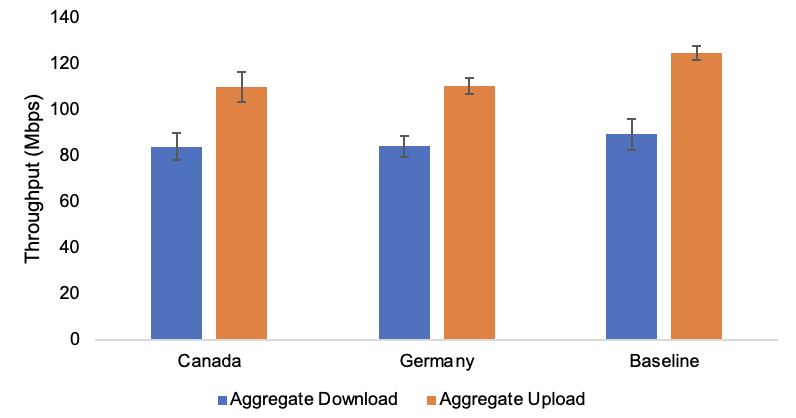}
    \caption{Throughput Comparison under 5 Clients. Error bars indicate $\pm\sigma$ over $5$ runs.}
    \label{fig:throughput_comparison}
   
\end{figure}

Table~\ref{tab:throughput_comparison} presents aggregate throughput, fairness metrics, and latency across client counts and configurations. As shown in Figure~\ref{fig:throughput_comparison}, the baseline scenario without VPN and proxy has a slightly higher throughput for both download and upload, revealing the modest overhead of our VPN/proxy stack ($\sim$5\% throughput reduction for download and $\sim$10\% for upload).

\paragraph{Bottleneck Identification}
Aggregate throughput remains nearly constant regardless of client count, demonstrating that the wireless link, not the proxy/VPN stack represents the primary bottleneck. Per-client throughput decreases as more devices connect, which results from shared and contention-based nature of the Wi-Fi (802.11) technology.

\paragraph{Bandwidth Distribution Fairness}
Jain's Fairness Index reveals near-optimal bandwidth distribution. Download fairness indices range from 0.95–0.99, indicating equitable sharing. Upload fairness is slightly lower, likely due to asymmetric traffic patterns. High fairness at 5 clients under maximum stress validates that ShieldShare does not introduce unfair allocation at the application layer.

\paragraph{Latency Stability}
Geographic distance introduces expected latency differences: Germany servers add ~150 ms compared to Canada. Critically, latency remains stable as client count increases. This stability indicates that proxy forwarding does not introduce queuing delays or processing bottlenecks under concurrent load.

\subsubsection{5 GHz Band Performance}
\label{subsubsec:5ghz_analysis}
Supplementary 5 GHz tests with 5 clients showed doubled throughput compared to 2.4 GHz (Canada VPN: 172.82 Mbps vs. 84.06 Mbps download). However, this higher throughput came with different fairness characteristics under stress conditions. Under stress conditions where all clients simultaneously attempted maximum throughput, the first-connected clients monopolized bandwidth in a queue-like pattern, contrasting with 2.4 GHz's fair distribution. This monopolization never occurred during realistic usage (browsing, streaming). We attribute this to 5 GHz offering more non-overlapping channels than 2.4 GHz, which reduces contention and increases throughput. The 2.4 GHz band provides predictable fairness through inherent contention, while 5 GHz offers higher throughput for staggered usage. While quota-based instead of throughput-based, ShieldShare's quota management (Section~\ref{subsubsec:quota}) can mitigate this monopolization.

\subsection{Resource Consumption}
\label{subsec:performance_monitoring}
To evaluate resource impact under realistic usage, we conducted 17-minute monitoring sessions with 1-4 clients performing mixed activities (YouTube streaming, web browsing). The Traffic Manager sampled battery level, CPU utilization, active connections, and throughput at 5-second intervals. This contrasts with the throughput stress tests where all clients simultaneously saturated bandwidth.

\subsubsection{Battery Consumption Analysis}
\begin{figure}[h]
    \centering
    \includegraphics[width=0.8\linewidth]{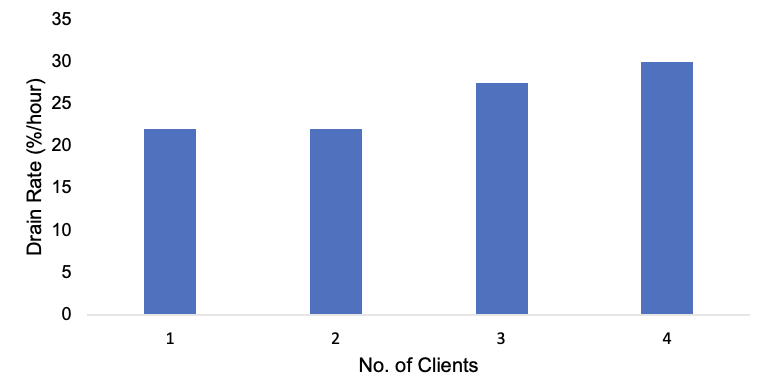}
    \caption{Battery Consumption Rates by Client Count}
    \label{fig:batteryconsumption}
    
\end{figure}

Battery drain represents the most significant operational constraint for extended sharing sessions. Figure~\ref{fig:batteryconsumption} summarizes battery impact across different client loads.

The results demonstrate that battery consumption increases as client count grows but at a decreasing rate, with diminishing marginal impact per additional client. Battery drain accelerates from approximately 22\%/hour with 2 clients to 30\%/hour with 4 clients. This progression indicates that the proxy/VPN stack's energy cost scales sub-linearly with connection count, though the absolute drain rate necessitates plugged-in operation for sessions exceeding 30 minutes with 3+ clients.

During the four-client extended test, the device exhibited thermal warnings after approximately 14 minutes of operation, suggesting that thermal management rather than CPU load becomes the limiting factor for sustained multi-client sharing on battery power.

\subsubsection{CPU Utilization Analysis}
\label{subsubsec:cpu_util}
CPU utilization was measured using process-specific tracking (\texttt{/proc/self/stat}), providing application-level metrics relative to a single CPU core. ShieldShare maintained an average CPU utilization of approximately 8-10\% across all tested scenarios, with occasional peaks reaching 65\% during connection establishment or high-throughput moments. These values reflect expected computational demands of VPN/proxy processing and indicate efficient operation within normal bounds for network workloads. The system maintains substantial headroom for additional features or higher client counts, with thermal constraints rather than CPU capacity representing the primary limiting factor for sustained multi-client sharing on battery power.

\section{Limitations}
While our proxy-based approach enables VPN sharing for hotspot clients, it introduces several inherent limitations.

\textbf{Reliance on Client-side Configuration:} Each client must configure their device to route traffic through the host's proxy server. This requirement arises because proxy server is not enabled by default when connected to a Wi-Fi. Consequently, only devices that allow proxy settings for Wi-Fi connections can utilize ShieldShare, excluding certain smart TVs, gaming consoles, and Internet of Things (IoT) devices.

\textbf{VPN Provider Policy Compliance:}
Beyond the technical compatibility issues discussed in Section~\ref{sec:vpn_compatibility}, some VPN providers may restrict connection sharing in their Terms of Service. Users should verify their provider's policies regarding connection sharing before using the app. ShieldShare does not circumvent provider restrictions; it operates within standard proxy and VPN protocols, but policy compliance remains the user's responsibility.

\textbf{No Stable Client Identifier:}
Clients are identified by DHCP-assigned IP addresses, which may change across sessions or reconnections. MAC address access which would enable persistent device tracking is restricted on Android without root privileges. As a result, traffic statistics and quota enforcement are session-based and do not persist across client reconnections.

\section{Conclusion}
ShieldShare successfully demonstrates a practical approach to secure network sharing on Android devices without requiring root access. By integrating a multi-protocol proxy server with VPN detection, per-client traffic metering, and traffic management, the system enables users to share their VPN-protected internet connection through a mobile hotspot while maintaining granular control over bandwidth allocation and usage monitoring. The implementation overcomes Android's permission restrictions through traffic-based client detection and provides a comprehensive solution for secure, measurable network sharing in collaborative environments. While the system faces limitations such as VPN compatibility constraints and performance overhead from concurrent monitoring, it establishes a foundation for future enhancements in cloud integration, enhanced security, and cross-platform optimization.

\bibliographystyle{ieeetr}
\bibliography{references}

\end{document}